\documentclass[prl,graphicx,reprint,twocolumn, superscriptaddress]{revtex4}
\usepackage{graphics,bm,overpic,subfigure,color}

\newcommand{\fsr}{\gamma_{E}}
\newcommand{\tprim}{R_0/L_T}

\newcommand{\gstwolnorm}{R_0} 
\newcommand{\gstworefspec}{i} 

\newcommand{\gstwoQnormflat}{n_\gstworefspec T_\gstworefspec v_{th\gstworefspec} \rho_{\gstworefspec}^2 /  \gstwolnorm^2   } 

\newcommand{\vct}[1]{\bm{#1}}

\newcommand{\lp}{\left(}
\newcommand{\rp}{\right)}

\newcommand{\rudolphpeierls}{Rudolph Peierls Centre for Theoretical Physics, University of Oxford, 1 Keble Road,  Oxford, OX1 3NP, UK}
\newcommand{\culham}{EURATOM/CCFE Fusion Association, Culham Science Centre, Abingdon, OX14 3DB, UK}
\newcommand{\isaacnewtoninstitute}{Isaac Newton Institute for Mathematical Sciences, 20 Clarkson Road, Cambridge, CB3 0EH, UK}

\newcommand{\edmundhighcock}{
\author{E.\ G.\ Highcock}
\email{edmund.highcock@physics.ox.ac.uk}
\affiliation{
\rudolphpeierls
}
\affiliation{
\culham
}
\affiliation{
\isaacnewtoninstitute
}

}

\newcommand{\michaelbarnes}{
\author{M.\ Barnes}
\affiliation{
\rudolphpeierls
}
\affiliation{
\isaacnewtoninstitute
}

}

\newcommand{\felixparra}{
\author{F.\ I.\ Parra}
\affiliation{
\rudolphpeierls
}
\affiliation{
\isaacnewtoninstitute
}

}

\newcommand{\colinroach}{
\author{C. M. Roach}
\affiliation{
\culham
}
\affiliation{
\isaacnewtoninstitute
}

}

\newcommand{\stevecowley}{
\author{S. C. Cowley}
\affiliation{
\culham
}
\affiliation{
\isaacnewtoninstitute
}

}

\newcommand{\alexschekochihin}{
\author{A.\ A.\ Schekochihin}
\affiliation{
\rudolphpeierls
}
\affiliation{
\isaacnewtoninstitute
}

}

\newcommand{\myfig}[2]{
\begin{figure}
\includegraphics{#1}%
\caption{#2\label{#1}}%
\end{figure}
}

\begin{document}


\title{Transport Bifurcation in a Rotating Tokamak Plasma} 

\edmundhighcock
\michaelbarnes
\alexschekochihin
\felixparra
\colinroach
\stevecowley

\date{\today}

\begin{abstract}
The effect of flow shear on turbulent transport in tokamaks is studied numerically in the experimentally relevant limit of zero magnetic shear. It is found that the plasma is linearly stable for all non-zero flow shear values, but that subcritical turbulence can be sustained nonlinearly at a wide range of temperature gradients. Flow shear increases the nonlinear temperature gradient threshold for turbulence but also increases the sensitivity of the heat flux to changes in the temperature gradient, except over a small range near the threshold where the sensitivity is decreased. A bifurcation in the equilibrium gradients is found: for a given input of heat, it is possible, by varying the applied torque, to trigger a transition to significantly higher temperature and flow gradients.
\end{abstract}


\maketitle 


\paragraph{Introduction.}

Turbulent transport of heat is a major obstacle to the development of a successful fusion device. Turbulence powered by microinstabilities such as the ion temperature gradient (ITG) instability rapidly transports heat out of the plasma, limiting the temperature gradient that can be sustained by a given input of heat, and thus the temperature that can be reached at the core of the plasma. The problem is exacerbated by the strong dependence of the turbulent amplitudes on the driving gradients, which in general keep the gradient not far above the critical threshold for the onset of the instability, a phenomenon known as stiff transport \cite{wolf2003internal}.

Experimental evidence \cite{burrell1997effects,connor2004itbreview} suggests that a sheared flow in the plasma greatly improves the situation; such a flow can significantly reduce turbulent fluxes for given values of the driving gradients. In some cases a large enough shear can help quench the turbulence altogether \cite{shafer2009internalq2,vries2009internal}. Flow shear may also reduce the sensitive dependence of heat flux upon temperature gradient (the "stiffness") \cite{manticastiffness}. Several of these results have been confirmed in numerical simulations \cite{waltz1998shear,   dimits2001parameter, kinsey2005flowshear, roach2009gss, casson2009anomalous}, but the picture remains incomplete.  Furthermore, while in simulations one may specify the flow shear, in experiment the control parameters are the input of heat and momentum. A set of simulations would ideally show not only that a large flow shear is beneficial, but also how it may be achieved.

A recent paper \cite{barnes2009transport} has demonstrated the basic properties of turbulence in the Cyclone Base Case \cite{dimits:969} regime (concentric circular flux surfaces with  \(q = 1.4\), \(r/R_0=0.18\), \(R_0\nu_{ii}/v_{thi}=0.01\), where \(q\) is the magnetic safety factor, \(r\) the half diameter of the flux surface measured in the midplane, \(R_0\) the major radius at the magnetic axis, \(v_{thi}\) the ion thermal velocity and \(\nu_{ii}\) the ion-ion collision frequency) at a finite value of magnetic shear. It was found that the ITG-driven turbulence was quenched at sufficiently high flow shear, but the ITG was replaced as the driver of the turbulence by the parallel velocity gradient (PVG) \cite{catto1973parallel, newton2010shearflow}. At large flow shears, the system was linearly stable, but strong subcritical PVG-driven turbulent transport could be sustained at sufficiently large temperature gradient.

In this Letter we consider the case where the shear of the magnetic field is zero, which experimental observations have indicated is favourable for quenching turbulence \cite{shafer2009internalq2,vries2009internal,yuh2009internal}. It is found that the plasma is now linearly stable for \emph{all} non-zero values of flow shear, and that there is a much larger range of flow shear and ITG values where the turbulence is completely quenched. It is shown that because of this (and when neoclassical transport is taken into account) there is a steady-state bifurcation in the flow and temperature gradients, for certain values of input heat and applied torque \cite{billbifurcationnote}. A positive feedback between the suppression of turbulence and the input of momentum can cause a one-way jump in both gradients.

\paragraph{Theoretical Framework.}

The work reported in this letter follows on from that carried out in \cite{barnes2009transport}, which contains a more detailed exposition of the model used. All simulations have been carried out using the code GS2 \cite{gs2ref}, which solves the local non-linear gyrokinetic equation in the presence of a sheared toroidal flow \(R\omega\) \cite{sugama1998neg, flowtome1}, where \(R\) is the major radius of the tokamak and \(\omega\) is the toroidal angular velocity. The flow is ordered as smaller than the ion thermal velocity in a Mach-number expansion: \(R\omega \sim M v_{thi}\) where \(\rho_i / R \ll M \ll 1\) and \(\rho_i\) is the ion Larmor radius. In order to ensure that the effect of the flow shear is retained, the gradient of the flow is then ordered as the inverse of the Mach number (\(d \ln\omega/d \ln r \sim 1 / M\)), so that the flow shear is of order the fluctuation frequency and the particle streaming rate: \(\fsr = (r/q)d\omega/dr\sim v_{thi} / R\). Using this expansion, the effects of the sheared toroidal flow are included in GS2 \cite{gs2flowshear, roach2009gss}, by adding a time dependence to the radial wavenumbers and a drive term associated with the PVG. All simulations reported here are electrostatic with a modified Boltzmann electron response. Typical resolution was \(32\times64\times14\times4\times8\) (poloidal, radial, parallel, pitch angle, energy). 

The gyrokinetic ordering is used to close the moment equations in the transport model \cite{sugama1997tpe, flowtome1}. The turbulent fluxes of heat and toroidal angular momentum, \(Q_t\) and \(\Pi_t\), respectively, can then be calculated from the perturbed ion distribution function \(\delta f\) as the flux surface averages of the convection of those quantities by the fluctuating \(\vct{E}\times \vct{B}\) velocity \cite{flowtome1}.

\paragraph{Subcritical Turbulence.}

With the flow shear \(\fsr\) equal to zero, the fluctuation amplitude grows exponentially with time (Fig. \ref{linear}). As the flow shear increases from zero, however, the growth becomes transient, and switches to decay after a time \(\tau_\gamma\) which decreases with increasing \(\fsr\) (inset to Fig. \ref{linear}).  This is in qualitative agreement with recent theoretical \cite{newton2010shearflow}, and numerical \cite{barnes2009transport} results at a finite value of magnetic shear, except that at zero magnetic shear we observe that there are no growing eigenmodes for any non-zero value of \(\gamma_E\).

\myfig{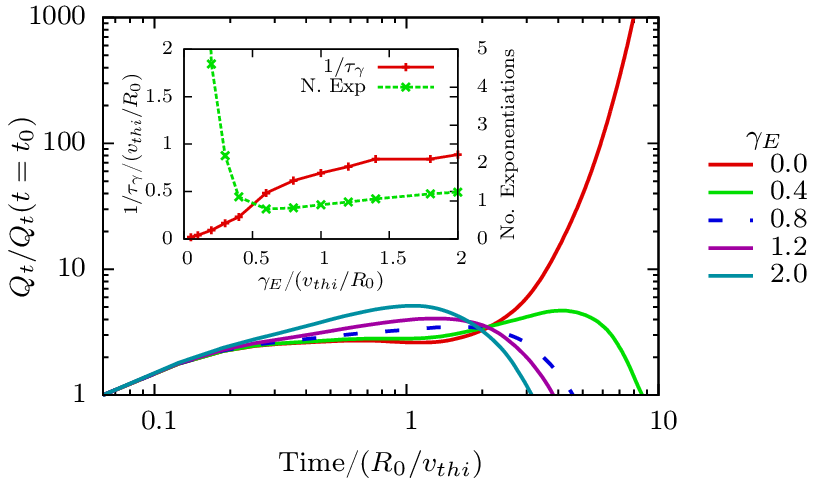}{Linear behaviour: the time evolution of the heat flux, normalised to its initial value, for different values of \(\fsr\) at \(\tprim=11\). Growth is transient for all non-zero flow shears, and the length of the period of transient growth decreases with flow shear. Dashed curve is subcritically stable (no turbulence sustained nonlinearly). Inset: Inverse transient growth time, and the number of exponentiations the heat flux undergoes before switching to decay, both vs. flow shear.}

With this in mind, it might be expected that there would be no turbulence for \(\fsr > 0\). However, linear instability is not necessary to sustain nonlinear turbulence in a rotating plasma; in fact, the transient growth caused by the ITG or PVG drive is sufficient to give rise to subcritical turbulence in simulations initialised with sufficient-amplitude noise (or fully developed turbulence). Thus, there is turbulence, but it is \emph{subcritical for all finite flow shears} (in contrast with the finite magnetic shear case \cite{barnes2009transport}).

\paragraph{Heat Flux.}

Considering the dependence of turbulent heat flux on the flow shear, Fig. \ref{fluxes}(a), we observe that for lower values of the temperature gradient, viz. \(R_0/L_T \lesssim 11.5\), where \(L_T\) is the temperature gradient scale length, \(Q_t\) decreases smoothly to zero with increasing flow shear. The turbulence is then fully quenched for a range of flow shears, but then the PVG drive becomes so large that turbulence is reignited. For larger values, \(R_0/L_T > 11.5\), the turbulence is never quenched; \(Q_t\) merely decreases and rises again, which echoes results in \cite{barnes2009transport}. We stress, however, that turbulence is quenched at lower values of flow shear and higher values of \(\tprim\) than in \cite{barnes2009transport}, and so there is a much wider range of parameter space where the flow shear is large enough to quench the ITG-driven turbulence, but not large enough to drive PVG turbulence. This more favourable regime is what enables the transport bifurcations described below.

\myfig{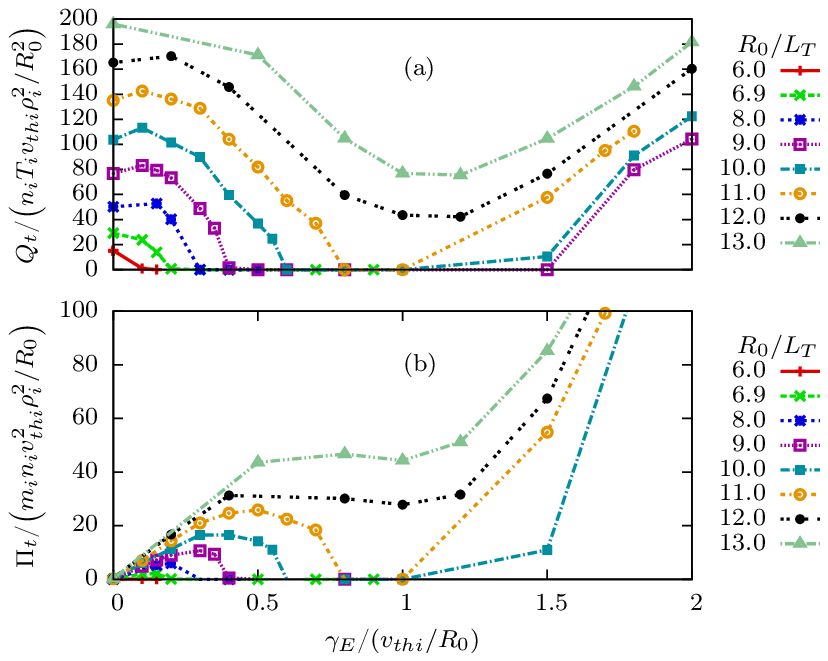}{Turbulent heat flux (a) and toroidal angular momentum flux (b) vs. flow shear for different values of \(\tprim\). For lower values of \(\tprim\), there is a range of flow shears where the turbulence is fully quenched.}

Examining the dependence of the heat flux on \(\tprim\), Fig. \ref{QvsTgrad} shows that at \(\fsr \leq 0.4\), there are two nonlinear thresholds (Fig. \ref{QvsTgrad}(c)), which both increase with flow shear. Below the first threshold turbulence is completely quenched; between the first and second thresholds \(Q_t\) increases slowly, and above the second it rapidly rises to values similar to the case without flow shear. Thus, flow shear significantly increases the overall temperature gradient required for turbulence, \emph{reduces} the transport stiffness dramatically between the first and second thresholds (i.e. at low values of \(Q_t\)), but \emph{increases} the stiffness above the second threshold (Fig. \ref{QvsTgrad}(d)). For \(\fsr \geq 0.8\) there is only one threshold, which decreases to 0 with increasing flow shear, (Fig. \ref{QvsTgrad}(b)), as the PVG drive increases. Stiffness is low but all the fluxes are very large.

\myfig{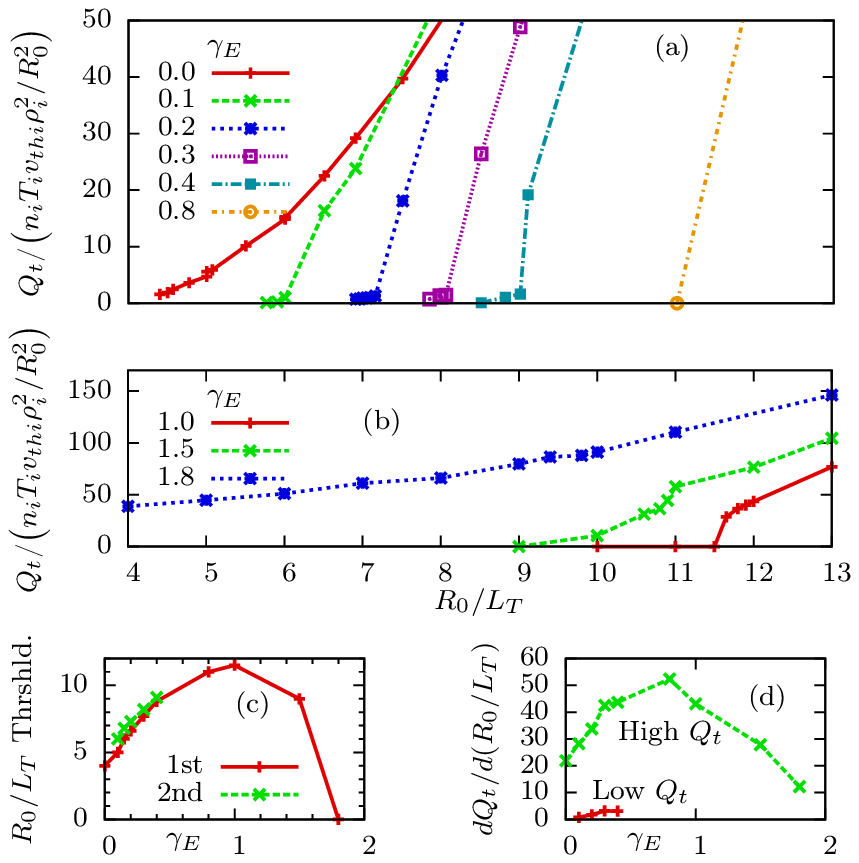}{Turbulent heat flux vs. \(\tprim\) for (a) \(\fsr \leq 0.8\) and (b) \(\fsr > 0.8\). (c) Nonlinear turbulence thresholds vs. flow shear. For each value of \(R_0/L_T\), the 1st-threshold curve shows the 2 values of \(\fsr\) at which the turbulence is quenched and then rekindled (cf. Fig \ref{fluxes}). For \(R_0/L_T \gtrsim 11.5\) or \(\fsr \gtrsim 1.8\), the turbulence is always present. (d) Profile stiffness vs. the flow shear. \emph{Low \(Q_t\)} refers to fluxes between the first and second thresholds, \emph{High \(Q_t\)} to fluxes above both thresholds.}

\paragraph{Momentum Flux.}

The story of the toroidal angular momentum flux \(\Pi_t\) is quite simple. Defining the turbulent viscosity as \(\nu_t = (\Pi_t / \fsr) (r / n_i m_i R_0^2 q) \), the turbulent heat diffusivity as \(\chi_t= \lp Q_t / \lp \tprim \rp \rp \lp  R_0 / (n_i T_i) \rp\) (where \(m_i\), \(n_i\) and \(T_i\) are the mass, density and temperature of the ions), the turbulent Prandtl number, \(Pr = \nu_t / \chi_t\), was in the range 1.0-1.8 in all our simulations, with only a very weak dependence on the gradients. In other words \(\Pi_t/Q_t \propto \fsr / (\tprim)\); the turbulence transports heat and momentum in equal proportions.

Fig. \ref{fluxes}(b) details this behaviour: \(\Pi_t\) rises with \(\gamma_E\), reaches a local maximum, drops, then rises again as the PVG starts to add significantly to the ITG turbulence drive. As with the heat flux, there is a window of zero turbulent transport for \(\tprim \lesssim 11.5\). The extent of this window in \(\fsr\) and \(\tprim\) is shown in Fig. \ref{QvsTgrad}(c).

\paragraph{Quenching Turbulence.}

At this point we abandon the use of \(\fsr\) and \(\tprim\) as independent variables, since the actual control parameters in a fusion device are the total rates of input of heat and momentum by external sources. If we assume a steady state, the rates of input of these quantities are equal to their outgoing fluxes, i.e. \(Q\) and \(\Pi\). In a practical manner, a question may be posed: in choosing \(Q\) and \(\Pi\), how can the temperature gradient, and hence the temperature at the core of the plasma, be maximised? Since the input of infinite amounts of heat is not possible, the turbulence must be quenched.

If turbulent transport is reduced, collisional transport becomes important. For tokamaks, a quantative theory of this transport, known as neoclassical theory, exists \cite{hintonwong1985nit}. For the case of circular concentric flux surfaces in the banana regime (\(\nu_{ii} \ll qR_0/v_{thi}\)) studied here, the neoclassical thermal diffusivity and viscosity are \(\chi_n\simeq 0.66 \lp R/r\rp^{\lp3/2\rp}q^2\rho_i^2 \nu_{ii}\)  and \(\nu_n \simeq 0.1 q^2 \rho_i^2 \nu_{ii}\). So the total heat flux \(Q = Q_t+Q_n\) where \(Q_n=\chi_n(\tprim) n_i T_i / R_0\), and the total toroidal angular momentum flux \(\Pi = \Pi_t+\Pi_n\) where \(\Pi_n=\nu_n\fsr n_i m_i R_0^2 q / r\).

The essential point is that the neoclassical Prandtl number, \(\nu_n/\chi_n \simeq  0.01\), is much smaller than the turbulent Prandtl number: turbulence is much more effective at transporting momentum than collisions alone. It should be noted that provided this is satisfied, the qualitative results of the next section do not depend on the exact values of the neoclassical transport coefficients.

\paragraph{Triggering a Transition.}

\myfig{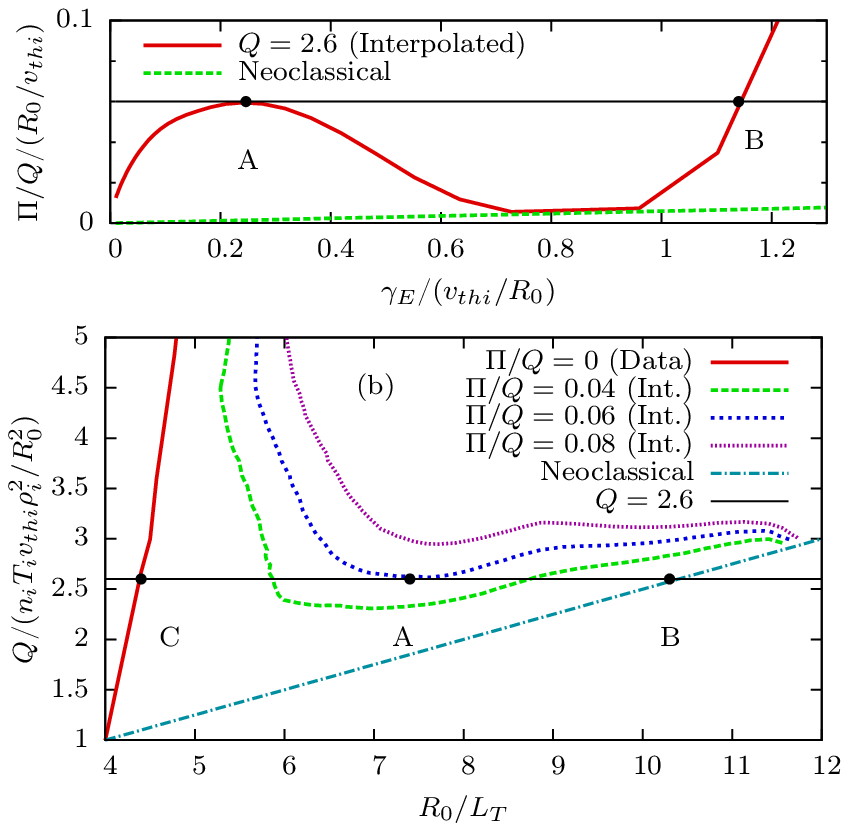}{(a) The ratio of the total momentum flux \(\Pi\) to the total heat flux \(Q\) vs. flow shear for a constant value of \(Q\). An increase in applied torque (i.e., an increase in \(\Pi/Q\)) at point A will cause a transition to point B.  (b) Total heat flux \(Q\) vs. \(\tprim\) for different constant values of \(\Pi/Q\). The points A and B on both graphs correspond to the same states. Since neither \(Q\) nor \(\Pi\) can be specified for a simulation (with the exception of \(\Pi = 0\)), the contours of constant \(Q\) and \(\Pi/Q\) were interpolated from a large number of data points using radial basis functions with a linear kernel \cite{buhmann2001radial}.  Also plotted are the neoclassical contributions to \(\Pi/Q\) and \(Q\). The contours in (b) do not intercept the neoclassical line but curl round and asymptote to it, tending back to the origin, and thus point B is in fact on the same contour as point A (see text and \cite{parra2010plausible}). This feature cannot be shown as the contours are too closely spaced for interpolation near the neoclassical line. }

Let us now consider what happens when we increase \(\Pi/Q\) at constant \(Q\). Fig. \ref{transition}(a) shows a curve where the heat flux is held constant at \(Q=2.6\) \(\gstwoQnormflat\). As \(\Pi/Q\) is increased, initially, because \(Q\) is being held constant, the suppression of the turbulence caused by an increase in \(\fsr\) causes a corresponding increase in \(\tprim\), which restores the turbulence to its former levels. However, at the point where \(\tprim\) becomes so large that \(Q_t \sim Q_n\), this negative feedback is broken, because \(Q_n\) does not depend on \(\fsr\). The turbulence is greatly reduced and yet the heat flux remains unchanged because the bulk of the heat is being transported neoclassically. The same is not true of the momentum, owing to the much lower neoclassical Prandtl number: because of the reduction in turbulent amplitudes, the transport of momentum drops dramatically. This causes a large increase in the flow shear gradient, which rekindles turbulence which then transports momentum once again. In other words, when the magnitude of \(\Pi/Q\) is increased above 0.06 \(R_0/v_{thi}\), there is a transition from point A to point B on Fig. \ref{transition}(a). 

Fig. \ref{transition}(b) shows the same transition on the (\(Q,\tprim\)) plane where contours of constant \(\Pi/Q\) are plotted. \(\fsr\) increases along these contours from high \(Q\) and low \(\tprim\) to low \(Q\) and high \(\tprim\). As \(\fsr\) increases, initially \(Q\) drops rapidly because of turbulence suppression by flow shear, until \(Q_n\) starts to become significant. At this point, because \(\Pi/Q \simeq \Pi_t/(Q_t + Q_n)\), \(\Pi_t\) must increase again to keep \(\Pi/Q\) constant. As flow shear increases further, however, \(\Pi_t/Q_t\) becomes so large (Fig. \ref{fluxes}) that the turbulence amplitude, and hence \(Q_t\), must start to decrease again to maintain a constant \(\Pi/Q\). Thus the curve asymptotes to the neoclassical line, and the system ends up in a state where \emph{the heat transport is nearly all neoclassical, but the momentum transport is nearly all turbulent}. Points A and B correspond to points A and B on \ref{transition}(a). Point B is not on the neoclassical line in Fig. \ref{transition}(b), but corresponds to a (slightly) turbulent state (as can be seen from Fig. \ref{transition}(a)), where the large \(\fsr\) means that \(\Pi_t\) is significant even though \(Q_t \ll Q_n\).  The location of point B on \ref{transition}(b) was calculated from \ref{transition}(a) using the relation \(\chi_t \sim \nu_t\). During the transition \(\tprim\) jumps from 7.4 to 10.4. Including the original suppression, flow shear has enabled a total jump (at constant \(Q\)) from \(\tprim \simeq 4.5\) at \(\Pi/Q=0\) (point C on Fig. \ref{transition}(b)), to \(\tprim\simeq10.4\) at \(\Pi/Q\simeq0.06\).

\paragraph{Conclusions.}

In summary, we have shown that although the plasma is linearly stable for all finite values of flow shear, transient growth is sufficient to allow subcritical turbulence to be sustained nonlinearly. At low values, flow shear reduces transport in two ways: by dramatically increasing the threshold temperature gradient required to drive turbulence, and, over a small range of temperature gradients, by reducing the strength of the dependence of the fluxes on the temperature gradient (the stiffness). At the top of that small range there is a second threshold, above which the fluxes rapidly rise to levels seen with no flow shear. High values of flow shear can in fact increase the transport.

Perhaps more importantly, we have discovered a transition to a higher-gradient regime. The transition occurs when either the input of heat has been reduced or the input of momentum increased to the point where the bulk of the heat is transported neoclassically and the ion temperature gradient is no longer driving sufficient turbulence to transport the toroidal angular momentum. A positive feedback loop starts where the build up of the velocity gradient reduces the turbulence further until it becomes sufficient to drive turbulent transport via the PVG instability. A new stable equilibrium is reached with much higher temperature and flow gradients, and where the heat transport is nearly neoclassical. 

The specific numbers associated with this transition are likely to depend significantly on the plasma configuration; for example, the value of \(q\) may have a large effect \cite{dimits2001parameter}. Extending this work to other regimes, in particular more experimentally realistic configurations, would thus be of interest. However, qualitatively we have shown not only that there exist equilibrium states with equal fluxes yet different gradients, but that, with the right conditions, less favourable regimes can automatically transition to more favourable ones with higher gradients.

\begin{acknowledgments}
We are grateful for helpful discussions with I. Abel, W. Dorland and G. W. Hammett. This work was supported by EPSRC (EGH, FIP), STFC (AAS) and the Leverhulme Network for Magnetised Plasma Turbulence. Computing time was provided by HPC-FF and by EPSRC grant EP/H002081/1. 
\end{acknowledgments}

\bibliography{references}

\end{document}